\documentclass[conference]{IEEEtran}
\IEEEoverridecommandlockouts
\usepackage{cite}
\usepackage{amsmath,amssymb,amsfonts}
\usepackage{algorithmic}
\usepackage{graphicx}
\usepackage{textcomp}
\usepackage{xcolor}
\usepackage{multirow}
\usepackage{booktabs}
\usepackage{tabularx}
\usepackage[ruled,linesnumbered]{algorithm2e}
\usepackage{fancyhdr}

\usepackage[letterpaper, margin=0.77in]{geometry}
\def\BibTeX{{\rm B\kern-.05em{\sc i\kern-.025em b}\kern-.08em
    T\kern-.1667em\lower.7ex\hbox{E}\kern-.125emX}}
\begin{document}

\title{UniMOS: A Universal Framework For Multi-Organ Segmentation Over Label-Constrained Datasets\\
\thanks{\textsuperscript{*} Corresponding author

\textsuperscript{\dag} These authors contributed equally to this work.
}
}

\author{\IEEEauthorblockN{1\textsuperscript{st} Can Li\textsuperscript{\dag}}
\IEEEauthorblockA{\textit{School of Cyber Science and Engineering} \\
\textit{Nanjing University of Science and Technology}\\
Nanjing, China \\
lican02@njust.edu.cn}
\and
\IEEEauthorblockN{1\textsuperscript{st} Sheng Shao\textsuperscript{\dag}}
\IEEEauthorblockA{\textit{School of Cyber Science and Engineering} \\
\textit{Nanjing University of Science and Technology}\\
Nanjing, China \\
shaosheng@njust.edu.cn}
\and
\IEEEauthorblockN{1\textsuperscript{st} Junyi Qu\textsuperscript{\dag}}
\IEEEauthorblockA{\textit{School of Cyber Science and Engineering} \\
\textit{Nanjing University of Science and Technology}\\
Nanjing, China \\
qujunyi@njust.edu.cn}
\and
\IEEEauthorblockN{2\textsuperscript{nd} Shuchao Pang\textsuperscript{*}}
\IEEEauthorblockA{\textit{School of Cyber Science and Engineering} \\
\textit{Nanjing University of Science and Technology}\\
Nanjing, China \\
pangshuchao@njust.edu.cn}
\and
\IEEEauthorblockN{3\textsuperscript{rd} Mehmet A. Orgun}
\IEEEauthorblockA{\textit{School of Computing} \\
\textit{Macquarie University}\\
Sydney, Australia \\
mehmet.orgun@mq.edu.au}
}
\maketitle
\thispagestyle{fancy}
\fancyhead{}
\lhead{}
\lfoot{978--8--3503--3748--8/23/\$31.00~\copyright~2023 IEEE}
\cfoot{}
\rfoot{}

\begin{abstract}
Machine learning models for medical images can help physicians diagnose and manage diseases. However, due to the fact that medical image annotation requires a great deal of manpower and expertise, as well as the fact that clinical departments perform image annotation based on task orientation, there is the problem of having fewer medical image annotation data with more unlabeled data and having many datasets that annotate only a single organ. In this paper, we present UniMOS, the first universal framework for achieving the utilization of fully and partially labeled images as well as unlabeled images. Specifically, we construct a Multi-Organ Segmentation (MOS) module over fully/partially labeled data as the basenet and designed a new target adaptive loss. Furthermore, we incorporate a semi-supervised training module that combines consistent regularization and pseudo-labeling techniques on unlabeled data, which significantly improves the segmentation of unlabeled data. Experiments show that the framework exhibits excellent performance in several medical image segmentation tasks compared to other advanced methods, and also significantly improves data utilization and reduces annotation cost. Code and models are available at: https://github.com/lw8807001/UniMOS.
\end{abstract}


\section{Introduction}
Multi-organ segmentation is an important component of medical image analysis and plays a crucial role in computer-aided diagnosis. For instance, accurate segmentation of at-risk organs in radiation therapy-based cancer and tumor treatment helps mitigate potential impacts on healthy organs near the cancerous region\cite{chen2021deep}. However, the medical image segmentation field faces challenges such as data scarcity and high annotation costs due to the difficulty of obtaining and annotating large-scale fully labeled datasets, which results in vast amounts of medical data left unannotated\cite{litjens2017survey}. Moreover, unlike other computer vision domains, most medical image datasets only collect segmentation for a specific type of an organ or tumor, with irrelevant organs and tumors labeled as the background. Therefore, another one of the key challenges in multi-organ segmentation tasks is how to utilize these partially annotated datasets for learning.

For addressing the issues of limited medical image annotation data and the phenomenon of many datasets only annotating a single organ, there are two mainstream strategies for improving image utilization. One strategy\cite{wu2022exploring,chen2023magicnet} is to leverage multi-category annotated images and utilize unlabeled images to accomplish multi-organ segmentation tasks. 
Another strategy\cite{zhang2021dodnet,wu2022tgnet,lian2023learning} is to use multiple partially annotated datasets to achieve multi-organ segmentation. Clearly, although these two strategies improve image utilization to some extent, they do not fully leverage the potential of all datasets. It would be meaningful and valuable if we could utilize all these datasets to train a multi-organ segmentation network in clinical applications.

\begin{figure*}[t!]
\centerline{\includegraphics[width=0.70\textwidth]{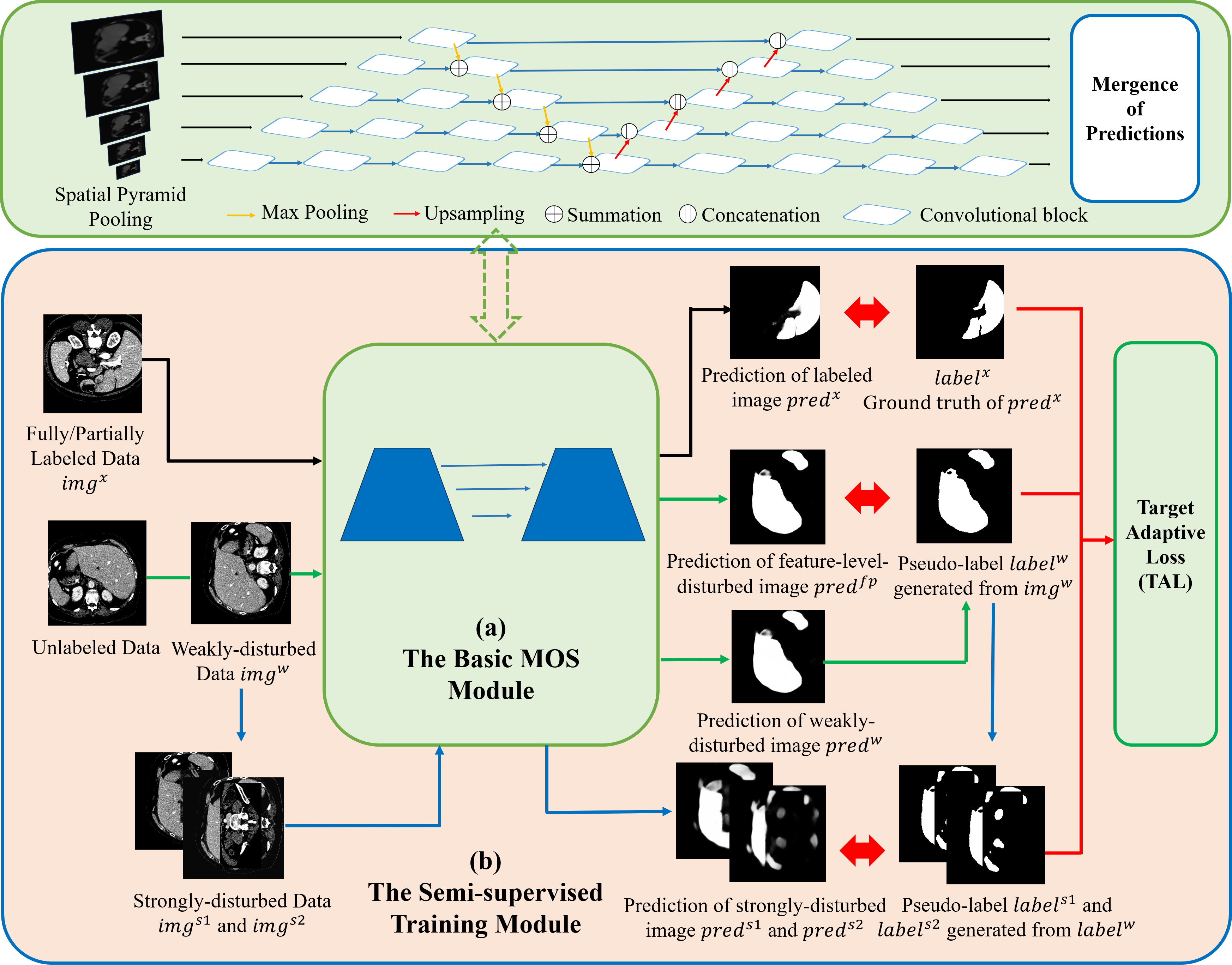}}
\caption{The architecture of UniMOS. UniMOS contains two modules, i.e., the basic MOS module and the semi-supervised training module. (a) The basic MOS module is capable of realizing multi-organ segmentation over various label-constrained data. (b) The semi-supervised training module makes full use of unlabeled data by utilizing a strong-to-weak consistency strategy.}
\label{model}
\end{figure*}

Overall, in order to address the above issues, we propose UniMOS, the first universal framework to fully exploit various label-constrained datasets, including fully and partially labeled images as well as unlabeled images in datasets. The simple architecture of UniMOS has been shown in Fig.\ref{model}.

The main contributions of this paper are as follows:

\begin{itemize}
    \item We propose a novel research topic with a new challenge from real clinical applications. Most of the previous multi-objective segmentation models can only be trained using datasets where all the categories are labeled at the same time, which results in poor model transferability.
    \item To address the new challenge and existing issues, we propose a universal learning framework, called UniMOS, to fully exploit these fully/partially annotated images and unlabeled images in the dataset and achieve image segmentation for multiple organs with a single model.
    \item We construct a multi-organ segmentation module that can be trained on multiple fully/partially labeled images at the same time and can integrate the results for simultaneous use in predicting organs across multiple categories. Moreover, we add a semi-supervised training module, which can be applied to a wider range of medical images, especially unlabeled images.
\end{itemize}

\section{Related Work}

\subsection{Semi-supervised Medical Image Segmentation}
Semi-supervised learning provides a method to improve model performance using unlabeled data. Currently, there are three typical paradigms in the field of semi-supervised medical image segmentation:

 \textbf{(a) Prior knowledge based approaches.} Medical images possess a significant amount of prior knowledge, such as the size and shape of organs. A common approach\cite{hu2021semi} is to use self-supervised pre-training, where medical image features of the same structure are learned and applied to downstream tasks.

\textbf{(b) Consistency regularization based methods.} In order to achieve more accurate segmentation results, numerous studies\cite{luo2022semi} have further optimized the perturbations at the input and feature map levels. Additionally, other methods\cite{zhao2023rcps} have focused on designing appropriate consistency regularization loss functions to balance consistency and accuracy in segmentation results.

\textbf{(c) Self-training based methods.} This kind of method involves assigning pseudo-labels to unlabeled images and then combining the pseudo-labeled images with the labeled images in order to update the segmentation model.  Many confidence or uncertainty-aware methods\cite{thompson2022pseudo} have been proposed in academia to generate more stable and reliable pseudo-labeling as a way to achieve more accurate semi-supervised medical image segmentation.

\subsection{Partially Labeled Medical Image Segmentation}
Medical image datasets usually contain only single or a few organs labeled because different institutes annotate the datasets for their own clinical research purposes. To solve the problem of not having large-scale fully labeled data, an intuitive strategy is to train a model on each partially labeled dataset, and then integrate the outputs from all models to achieve the multi-organ segmentation task. However, this greatly increases the computational complexity and storage overhead. 

Currently, one advanced and major strategy is to train a unified model with multiple partially labeled datasets, but similar research is still very limited. To realize this strategy, some scholars\cite{zhang2021dodnet, wu2022tgnet} have proposed new learning framework or new learning method. In addition, some scholars\cite{lian2023learning} have attempted to utilize prior knowledge to guide learning.

\section{Methods}

\subsection{The Basic MOS Module}

Inspired by PIPO-FAN\cite{fang2020multi}, i.e., pyramid input and pyramid output feature abstraction  network, our proposed UniMOS framework also applies spatial pyramid pooling and shared convolution to input data in order to get contexts in different scales. Its multi-scale features at each depth combine global and local multi-scale contexts into a single model to enhance the extracted features.

At the same time, we resort to its target adaptive loss (TAL) as loss function to utilize partially labeled data, i.e. merging background predictions and category predictions not included in the currently used data set as:

\begin{equation}
L_{TAL}=\sum_{c\in C_{k}}y^{c}_{i}log p^{c}_{i}+ \mathbb{I}_{[\sum_{c\in C_{k}}y^{c}_{i}=0]}log (1-\sum_{c\in C_{k}} p^{c}_{i}),\label{eq1}
\end{equation}
where $c$ is the class of current voxel $i$, $C_{k}$ are classes included in data set used in current epoch. $p^{c}_{i}$ is the prediction of voxel $i$ belonging to class $c$, $y^{c}_{i}$ means whether voxel $i$ belonging to class $c$ or not.

By TAL, UniMOS regards voxels belonging to target class as foreground and other voxels as background, converting multi-class segmentation to binary segmentation,  which makes it possible to train a model for multi-organ segmentation using partially labeled data. 

\subsection{The Semi-supervised Training Module}
By leveraging a semi-supervised training strategy, UniMatch\cite{yang2023revisiting} which improved the classical semi-supervised training method FixMatch\cite{sohn2020fixmatch}, the proposed UniMOS framework also extend it into our semi-supervised training module. Specifically, it is based on the idea of consistent regularization and pseudo-labeling for semi-supervised training and applies image-level or feature-level disturbances to improve the segmentation performance of unlabeled data.

\begin{figure}[h]
\centerline{\includegraphics[width=0.5\textwidth]{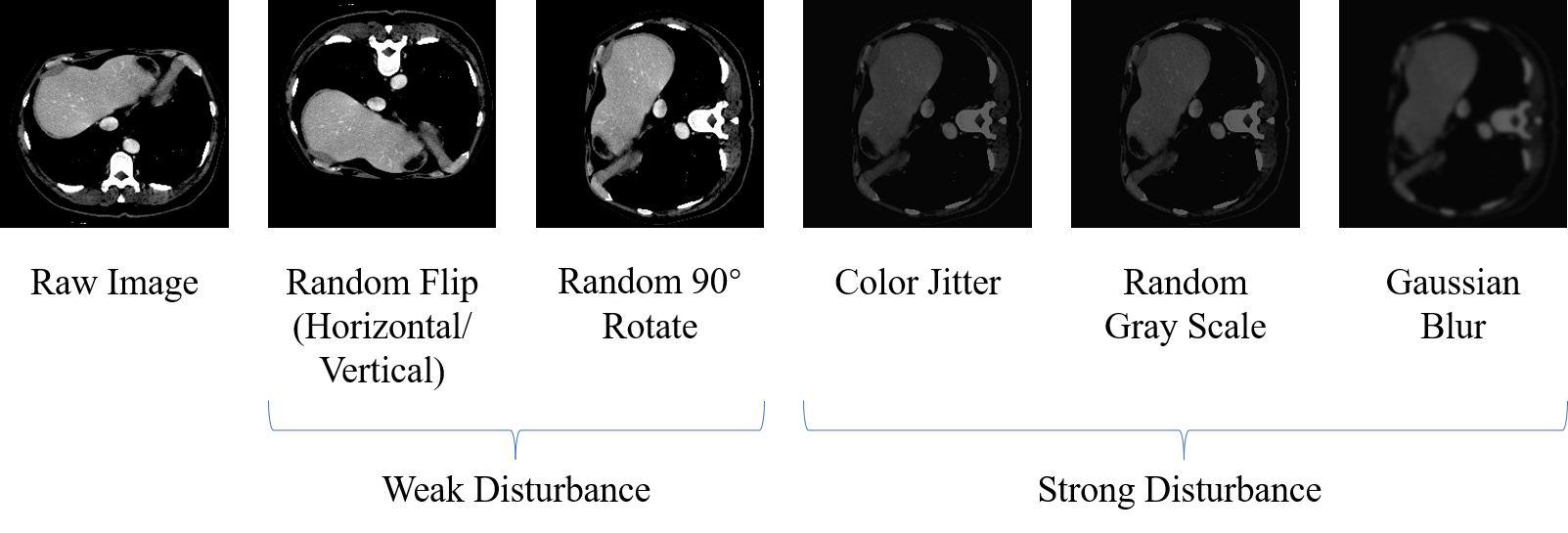}}
\caption{Weak disturbances and strong disturbances used in UniMOS. Applied in left-to-right order.}
\label{fig2}
\end{figure}

\textbf{Image-level disturbances}. For the labeled data $x^l$ , we input it directly into the MOS module and obtain the prediction $p^l$, and calculate with the ground truth $y^l$ to get the supervised loss $L^l$.
For unlabeled data, we apply a weak disturbance (random horizontal/vertical flip, random 90° rotation) to the original data to obtain the weakly-disturbed version $x^w$ used for generating pseudo-labels $y^w$. Then strong disturbances (color jitter, random gray scale, Gaussian blur, Cutmix) as shown in Fig.\ref{fig2} are applied to $x^w$ to obtain the strong-disturbed version $x^s$.

\textbf{Feature-level disturbances}. We also disturb the features from the encoder to obtain the feature-level disturbed data. UniMOS uses the same feature-level disturbance as UniMatch, i.e., applying nn.Dropout2d(0.5) to the features.

\textbf{Dual-stream disturbances}. UniMOS uses two strongly-disturbed versions of $x^w$, $x^{s1}$ and $x^{s2}$, to train the MOS module. This dual-stream disturbance approach can further improve the segmentation performance of UniMOS.

\textbf{Generate pseudo-labels}. UniMOS generates pseudo-labels from $x^w$’s prediction $p^w$. We use the class corresponding to the maximum prediction as the pseudo-label and the maximum prediction as the confidence level. Only pixels with confidence higher than a threshold $\tau$ are used for loss.  

\textbf{Loss function}. Predictions of feature-level-disturbed data and strongly-disturbed data will be calculated with pseudo-labels together to figure out the semi-supervised loss $L^u$ as: 

\begin{equation}
L^{s} = \frac{1}{N}\sum(f(p^{s}, y^w))\mathbb{I}(max(p^{s1})>=\tau),\label{eq2}
\end{equation}
\begin{equation}
L^{fp} = \frac{1}{N}\sum(f(p^{fp}, y^w))\mathbb{I}(max(p^{fp})>=\tau),\label{eq4}
\end{equation}
\begin{equation}
L^{u} = 0.25 * L^{s1} + 0.25 * L^{s2} + 0.5 * L^{fp},\label{eq5}
\end{equation}
where $N$ is batch size and $f$ is loss function. 

The supervised loss $L^l$ and semi-supervised loss $L^u$ are averaged to obtain the final loss as:
\begin{equation}
L = 0.5 * L^{l} + 0.5 * L^{u}.\label{eq6}
\end{equation}

\begin{figure}[t!]
\centerline{\includegraphics[width=0.43\textwidth]{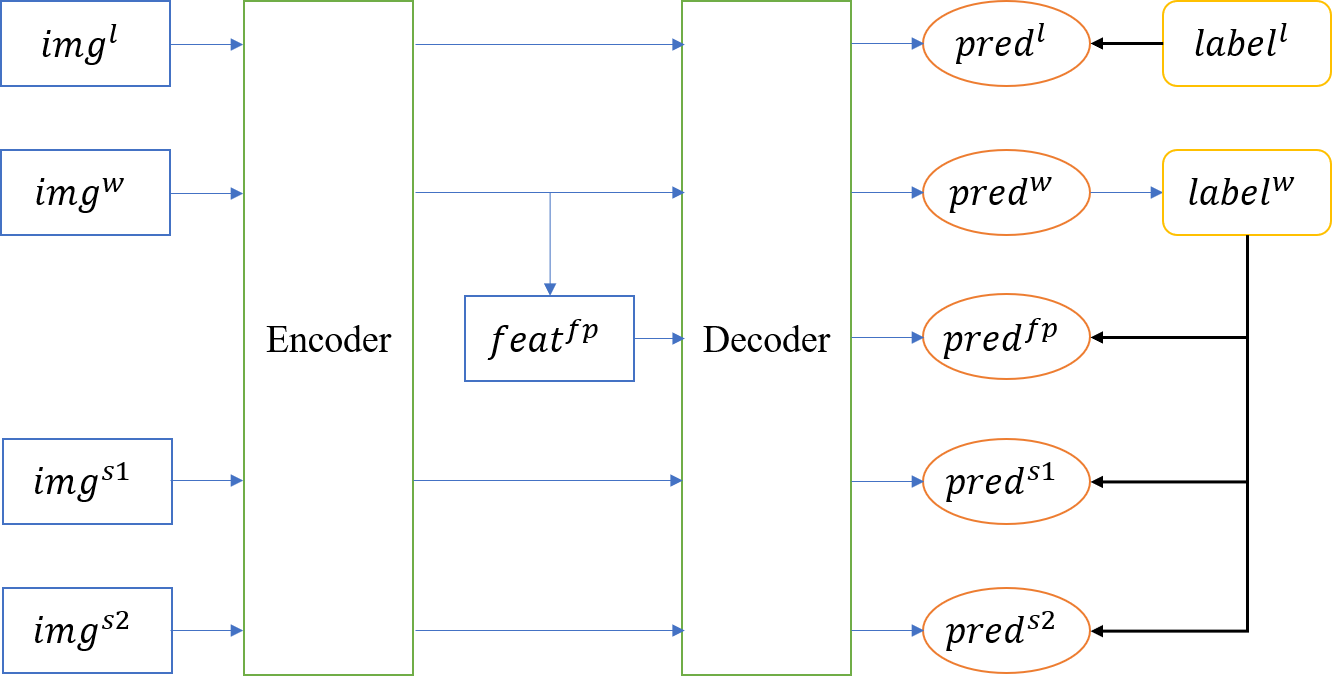}}
\caption{UniMOS's training process. $img$ means "image", $feat$ means "feature of $img^w$ generated from encoder", $pred$ means "prediction of image". $l$ means "labeled data", $w$ means "weakly-disturbed data", $s1$ and $s2$ mean "differently strongly-disturbed data".}
\label{fig3}
\end{figure}

The data used in UniMOS and its training process are shown in Fig.\ref{fig3}.

\section{Experiments}
\subsection{Setup}
\textbf{Datasets}. We evaluated our UniMOS framework on several publicly available data sets including LiTS (Liver Tumor Segmentation Challenge), KiTS (Kidney Tumor Segmentation Challenge) and MSDSpleen (Medical Segmentation Decathlon Spleen). They were used as partially labeled data sets for training. 

\textbf{Implementation details}. UniMOS trained $2500$ epochs with initial learning rate $0.0005$ which decays by $0.01$ after every $40$ epochs. We use RMSprop as optimizer. The data need to be resized to $256*256$ initially, whose resolution will be $224*224$ after MOS module pre-processing. In addition, we used Dice score as the criterion to evaluate the segmentation performance of UniMOS.

\begin{table}[t!]
\caption{Dice Score on Each Organ and speed from Different Methods. Note that Tr. is the training time and Te. is the testing time}
\resizebox{1\columnwidth}{!}{
\begin{tabular}{cllllcc}
\hline
\multicolumn{1}{l}{\multirow{2}{*}{}} & \multicolumn{4}{c}{Dice Score(\%)}    & \multicolumn{2}{c}{Speed} \\ \cline{2-7} 
\multicolumn{1}{l}{} &
  \multicolumn{1}{c}{Liver} &
  \multicolumn{1}{c}{Kidney} &
  \multicolumn{1}{c}{Spleen} &
  \multicolumn{1}{c}{Average} &
  \multicolumn{1}{l}{Tr.(min/epoch)} &
  \multicolumn{1}{l}{Ts.(s/case)} \\ \hline
UniMOS  & \textbf{91.67\%} & 86.20\%    & \textbf{93.10\%} & \textbf{90.32\%} & 2.15    &  8.43  \\
KiU-Net\cite{valanarasu2020kiu}   & 83.94\%    & 77.56\%     & 75.36\%     & 78.95\%     & \textbf{0.10}    & 8.57    \\
DoDNet\cite{zhang2021dodnet}     & 90.94\%     & \textbf{88.04\%} & 89.31\%    & 89.43\%        & 8.56       & 54.55      \\ 
PIPO-FAN\cite{fang2020multi}   & 86.62\%    & 78.32 \%     & 88.34\%     & 84.43\%     & 0.62    & \textbf{8.16}    \\ \hline
\end{tabular}
}
\label{tab2}
\end{table}

\subsection{Comparison with Other Methods}

We compared UniMOS with other advanced methods, including those attempt to utilize multiple partially-labeled data sets (e.g., DoDNet\cite{zhang2021dodnet}, PIPO-FAN\cite{fang2020multi}) and those use traditional fully-labeled data sets to train multiple models for different tasks (e.g., KiU-Net\cite{valanarasu2020kiu}).

For a fair comparison, we removed the offline post-processing step applied to the prediction from original methods\cite{zhang2021dodnet,valanarasu2020kiu} because UniMOS didn’t employ post-processing technique.

The experimental results show that:
\begin{itemize}
\item By comparing the results of PIPO-FAN, we conclude that using more data (even unlabeled data) improves the performance of the model. With the help of rich labeled restricted data, UniMOS achieves excellent performance, with the average Dice of its prediction results exceeding those of the methods we compare.

\item The training process of DoDNet is quite slow because of the large computational cost and the complicated iterative process to ensure the training accuracy. UniMOS has a faster training speed with guaranteed performance. 

\end{itemize}

\section{Conclusion}
To fully utilize all types of label-constrained data in one model, we propose the universal framework, called UniMOS, for multi-organ segmentation tasks, including a basic MOS module and a semi-supervised training module. Furthermore, UniMOS leverages different kinds of fully and partially labeled data as well as unlabeled data for semi-supervised training to achieve the segmentation of multiple organs with a single model.

UniMOS can reduce the cost of data annotation which has a high application potential.
In the future, we conjecture that UniMOS can be used for image segmentation on non-medical data sets after slightly changing the functions of the data loader and pre-processing.

\section*{Acknowledgments}

This work is supported by the National Natural Science Foundation of China under Grants No. 62206128, the Undergraduate Research Training Program of Nanjing University of Science and Technology under Grants No. 2022066017B and the Postgraduate Research \& Practice Innovation Program of Jiangsu Province No.SJCX23\_0115. The authors declare no conflict of interest.

\bibliographystyle{IEEEtran}
\bibliography{references}
\end{document}